\documentclass[acmtog]{acmart}
\acmSubmissionID{614}
\AtBeginDocument{%
  }

\definecolor{lghcol}{rgb}{0.2,0.6,0.2}

\newcommand{\ep}[1]{{\color{black} #1}}

\setcopyright{acmlicensed}

\copyrightyear{2025}
\acmYear{2025}
\setcopyright{acmlicensed}\acmConference[SIGGRAPH Conference Papers '25]{Special Interest Group on Computer Graphics and Interactive Techniques Conference Conference Papers }{August 10--14, 2025}{Vancouver, BC, Canada}
\acmBooktitle{Special Interest Group on Computer Graphics and Interactive Techniques Conference Conference Papers (SIGGRAPH Conference Papers '25), August 10--14, 2025, Vancouver, BC, Canada}
\acmDOI{10.1145/3721238.3730672}
\acmISBN{979-8-4007-1540-2/2025/08}

\begin{document}

%%
%% The "title" command has an optional parameter,
%% allowing the author to define a "short title" to be used in page headers.
\title{Model See Model Do: Speech-Driven Facial Animation with Style Control}

%Example-based expressive speech style transfer with style basis
%Performance by Example: Stylized Facial Animation from Speech

%%
%% The "author" command and its associated commands are used to define
%% the authors and their affiliations.
%% Of note is the shared affiliation of the first two authors, and the
%% "authornote" and "authornotemark" commands
%% used to denote shared contribution to the research.
\author{Yifang Pan}
% \authornote{Both authors contributed equally to this research.}
\email{evan.pan@mail.utoronto.ca}
\orcid{0000-0001-5194-0359}
\affiliation{%
  \institution{University of Toronto}
  \city{Toronto}
  \state{Ontario}
  \country{Canada}
}

\author{Karan Singh}
\email{karan@dgp.toronto.edu}
\affiliation{%
  \institution{University of Toronto}
  \city{Toronto}
  \state{Ontario}
  \country{Canada}}
  
\author{Luiz Gustavo Hafemann}
\email{luiz.gustavo-hafemann@ubisoft.com}
\affiliation{%
  \institution{Ubisoft La Forge}
  \city{Montreal}
  \state{Quebec}
  \country{Canada}
}

%%
%% By default, the full list of authors will be used in the page
%% headers. Often, this list is too long, and will overlap
%% other information printed in the page headers. This command allows
%% the author to define a more concise list
%% of authors' names for this purpose.
% \renewcommand{\shortauthors}{Trovato et al.}

%%
%% The abstract is a short summary of the work to be presented in the
%% article.
\begin{abstract}
Speech-driven 3D facial animation plays a key role in applications such as virtual avatars, gaming, and digital content creation. While existing methods have made significant progress in achieving accurate lip synchronization and generating basic emotional expressions, they often struggle to capture and effectively transfer nuanced performance styles. We propose a novel example-based generation framework that conditions a latent diffusion model on a reference style clip to produce highly expressive and temporally coherent facial animations. To address the challenge of accurately adhering to the style reference, we introduce a novel conditioning mechanism called style basis, which extracts key poses from the reference and additively guides the diffusion generation process to fit the style without compromising lip synchronization quality. This approach enables the model to capture subtle stylistic cues while ensuring that the generated animations align closely with the input speech. Extensive qualitative, quantitative, and perceptual evaluations demonstrate the effectiveness of our method in faithfully reproducing the desired style while achieving superior lip synchronization across various speech scenarios.

\end{abstract}

%%
%% The code below is generated by the tool at http://dl.acm.org/ccs.cfm.
%% Please copy and paste the code instead of the example below.
%%

% \ccsdesc{\TODO{todo}}
% \ccsdesc[100]{Do Not Use This Code~Generate the Correct Terms for Your Paper}

%%
%% Keywords. The author(s) should pick words that accurately describe
%% the work being presented. Separate the keywords with commas.
%% A "teaser" image appears between the author and affiliation
%% information and the body of the document, and typically spans the
%% page.
\begin{teaserfigure}
  \includegraphics[width=\textwidth]{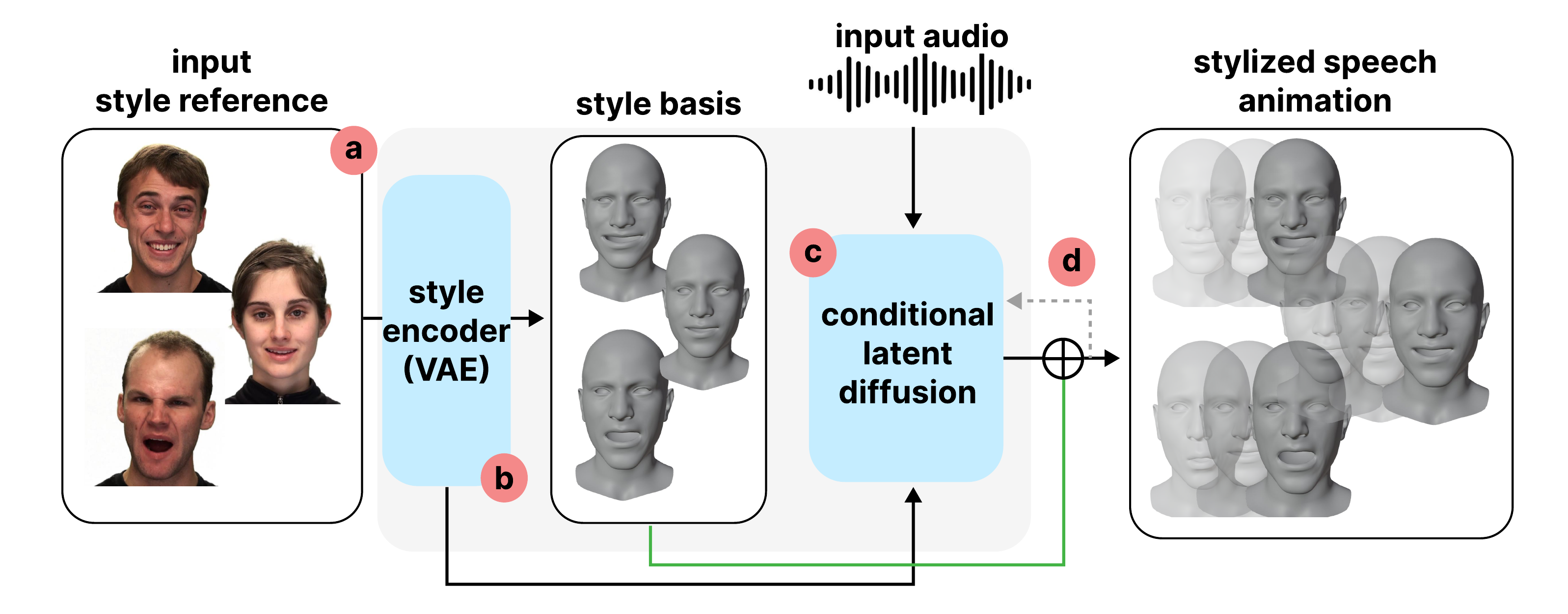}
  % \rule{\textwidth}{6cm}
  \caption{
  %We present an example-based system for generating stylistic 3D facial animations. The pipeline consists of the following steps: (a) The model takes an arbitrary motion sequence as a reference for the output motion, (b) which is encoded into a latent style feature that captures the performance style and a style basis that reflects key poses from the reference, (c) a conditional diffusion module, conditioned on both the audio and the latent style feature, generates a prototype motion, and (d) the prototype motion is recombined with the style basis to produce the final output animation.
  We present an example-based system for generating stylistic 3D facial animations: (a) Given an arbitrary style reference, a style encoder (b) obtains latent style features and a style basis that reflects key poses from the reference. (c) A diffusion module, conditioned on audio and style features, produces the primary motion. (d) The style basis guide the primary motion throughout the diffusion process, gradually refining it at each diffusion step to produce the final animation.
  }
  \label{fig:teaser}
\end{teaserfigure}

% \received{9 January 2025}
% \received[accepted]{13 April 2025}

%%
%% This command processes the author and affiliation and title
%% information and builds the first part of the formatted document.
\maketitle

\section{Introduction}
% \plan{Speech-driven animation is cool and useful}

%Automatically generating realistic facial animation from speech has long been a compelling yet challenging goal in the fields of computer graphics and human–computer interaction.
Automatically generating realistic 3D facial animations from speech has long been a compelling yet challenging goal in computer graphics, with applications in film, virtual reality, video games, and education. 

% \plan{Current literature addresses lip-sync, but not much work on controlling generation. Controlling the style is an important problem}

%While recent deep generative methods produce increasingly convincing lip-sync motions [Sun, 2024][Xu, 2024][Zhao, 2024], they often lack an intuitive method of authoring the style of the performance.

Recent deep generative methods produce realistic facial movements with accurate lip-sync \cite{xing_codetalker_2023,sun_diffposetalk_2023, zhao_media2face_2024}. These methods use datasets of paired audio and sequences of 3D meshes, and learn to generate a sequence of facial movements that match the speech. While good lip synchronization may suffice for some applications, \emph{how} the speech is conveyed is equally important for generating compelling facial performances. There are many ways the same line can be delivered, as many facial movements are only weakly correlated with audio (e.g. movements of eyebrows, forehead, range of the mouth articulation). Providing an intuitive method for \emph{controlling} the motion generation remains an open problem.

%
%Most methods in the literature focus on they often lack an intuitive method of authoring the style of the performance.

% \plan{Existing methods for controllability are flawed.}

Some recent work attempts to add control signals for the generation, by using emotion labels \cite{danecek_emotional_2023} and text \cite{zhao_media2face_2024}. In practice, speech performance is influenced by myriad factors (e.g., character quirks, emotional tone, or speaker identity), making it difficult to annotate or discretize these dimensions without imposing rigid categories. A more flexible approach is to condition the generation based on examples, as investigated in domains such as speech generation \cite{zaidi_daft-exprt_2022, wang_style_2018} and co-speech gesture generation \cite{ghorbani_zeroeggs_2022}. 

\cite{sun_diffposetalk_2023} provides a first attempt at adding example-based control for speech-driven facial animation. In their work, style is learned in a separate step, with a contrastive loss. However, their approach has key limitations. By using a contrastive loss in a separate training step, the style encoding is learned independently from the main generation task. This means the style encoder cannot learn which facial movements are determined by speech content and which are stylistic choices that vary between performances. Moreover, the contrastive learning framework treats all non-matching examples as equally dissimilar negatives, failing to capture the natural hierarchy of stylistic similarities - for instance, two \emph{angry} performances would be pushed as far apart as an \emph{angry} and a \emph{happy} performance.

%Existing controllability mechanisms, such as text prompts or emotion labels—tend to be coarse, expensive to collect, and insufficient for capturing the richness of expressive or personalized speech styles. In practice, speech performance is influenced by myriad factors (e.g., character quirks, emotional tone, or speaker identity), making it difficult to annotate or discretize these dimensions without imposing rigid categories.

% \plan{We propose a control/stylization method to address these flaws}

%We propose a new \emph{example-based} approach for style control, that addresses these limitations. Rather than requiring detailed text descriptions or discrete emotion tags, we take a short reference motion clip and leverage it to convey a broad range of stylistic cues: from subtle upper-face expressions to characteristic lip-shapes. Such example-based control offers a more intuitive user interface—people can simply select a performance clip that “looks right”—and it encodes fine-grained variability unattainable through traditional categorical labels [Ghorbani, 2022][Valle-Pérez, 2021]. Moreover, referencing personal style clips enables easy adoption of new speakers or emotional nuances without massive re-labeling effort.

We propose a new \emph{example-based} approach for style control that addresses these limitations. Our method takes an audio signal and a style reference, and generates facial animation matching both.
The core insight underlying our method is that expressive speech motion can be decomposed into two complementary components: 
1) \emph{Time-invariant expressions}, which capture a speaker's overarching expressions throughout line delivery (e.g., squinting eyes or furrowed brows), and
2) \emph{Motion dynamics}, which corresponds to the movements directly tied to phonetic content, capturing how a speaker articulates words. We incorporate this decomposition into a \emph{latent diffusion} model, where the time-invariant expressions modulate the motion dynamics iteratively throughout the diffusion process. 
By jointly training the style encoder with the motion diffusion model, our method is better suited to extract the nuances of the style reference, and use it to generate facial motion that matches both the style and audio content.

We conducted experiments using a large-scale dataset of paired audio and facial motion data. 
% To this end, we designed a thorough procedure to curate existing in-the-wild video datasets, and used a state-of-the-art capture method to extract facial animations. 
% We compare our model to the state-of-the-art stylization models of \cite{sun_diffposetalk_2023} and \cite{ghorbani_zeroeggs_2022}, using a strict experimental protocol,
We compare our model to the state-of-the-art stylization model of \cite{sun_diffposetalk_2023}  using a strict experimental protocol, where only the style encoding mechanism is varied. We measured standard metrics on lip synchronization and upper-face movements. We also conducted two user studies to measure the ability of the models in maintaining the style of reference clips, and obtaining accurate lip sync. Our model shows improvement in lip synchronization, upper-face dynamics and style transfer, both in quantitative tests and user studies.
%By jointly training a variational auto-encoder (to encode the example video as a single style vector) and a conditional diffusion pipeline (to generate both the time-invariant and time-varying motion), our system not only achieves strong quantitative performance on benchmark evaluations but also produces animations that more faithfully reflect the subtle style of the example clip, confirmed by User studies.

%\epfix{There are some things missing here. We should briefly mention about the other stuff we did: data curation, and the user study design to separately study the lip sync and style transfer}

% \plan{We also acknowledge the lack of stylized datasets and provide tools to curate a dataset.}

In summary, our key contributions include:
\begin{itemize}
    % \item \epfix {we proposed example driven facial animation that captures nuances and acutally works.} 
    \item An example-based speech-driven facial animation framework that captures the speech mannerisms of the reference style, with robust user studies validating its effectiveness in accurately preserving style nuances.
    \item A novel conditioning mechanism for diffusion-based facial animation generation, called style basis, experimentally shown to improve motion accuracy and expressiveness through ablation studies.  
    \item We release our model's training code and data curation pipeline, to facilitate future research using in-the-wild videos for training and evaluation.

\end{itemize}

%\epfix{We need to think how to improve this - these are easy targets for a reviewer to criticize: #1 and #2 are kind of describing the same thing; it can be argued that 2 already introduce the paradigm, just didn't have good results. }

\section{Related Works}
In these sections, we will discuss related works in three relevant fields, including automatic lip-sync animation, style-conditioned generative models, and latent diffusion models.  

\subsection{Automatic 3D Lip-sync Animations}
The study of automatic 3D lip-sync animation spans several decades, with the correspondence between phonemes (vocalized sounds) and visemes (lip shapes) discovered as early as 1968 \cite{fisher_confusions_1968}. Early lip-sync systems were often procedural, using phoneme alignment to obtain the sequence of phoneme timings, then generate the corresponding sequence of visemes \cite{edwards_jali_2016, cohen_modeling_1993, pan_vocal_2022, bailly_animated_2012}.
%LGH: needs some work

More recently, deep learning-based auto-regressive models have been shown to successfully capture 3D lip shapes without explicitly modeling phonemes and visemes \cite{xing_codetalker_2023, richard_meshtalk_2022, karras_audio-driven_2017, fan_faceformer_2022}. However, due to the lack of publicly available 3D speech datasets with rich emotions, these models often focus on relatively neutral speech from the datasets BIWI \cite{fanelli_3-d_2010} and VOCASET \cite{cudeiro_capture_2019}.

Instead of using small datasets of high-quality 3D facial capture, other works build larger-scale datasets from in-the-wild videos. They rely on advancements in monocular facial capture methods \cite{feng_learning_2021, danecek_emoca_2022, josi_serep_2024} to extract 3D meshes from videos, commonly modeled as coefficients of a 3D Morphable Model \cite{egger_3d_2020}. While the quality of the 3D capture from these methods is inferior to that of BIWI and VOCASET, this enables processing large-scale video datasets such as CelebV-text \cite{yu_celebv-text_2023} and CelebV-HQ \cite{zhu_celebv-hq_2022}, and expressive video datasets such as RAVDESS \cite{livingstone_ryerson_2018} and MEAD \cite{vedaldi_mead_2020}. These datasets have enabled the training of generative models conditioned on emotional categorical labels, such as EMOTE \cite{danvevcek2023emotional} and EmoTalk \cite{peng_emotalk_2023}, and text-conditioned models such as Media2Face \cite{zhao_media2face_2024}.
%\epfix{Upon visual examination, we observed that SEREP captures significantly more expressive facial motion compared to the Mead3D dataset processed with EMOCA \cite{danecek_emoca_2022} and the HDTF dataset processed by \cite{sun_diffposetalk_2023}.}

% \epfix{Check for other references to [Livingstone, 2018] - that's the wrong "EMOTE" paper. The correct one is from danecek}

%With advancements in monocular facial capture [Danecek, 2022] [Feng, 2021][Josi, 2024], it is now possible to extract 3D lip-sync data from labeled and emotionally expressive audiovisual datasets such as RAVDESS [Livingstone, 2018] and MEAD [Wang, 2020]. These datasets have enabled the training of generative models conditioned on emotional categorical labels, such as Emote [Senadeera, 2023] and Emotalk [Peng, 2023].

%Similarly, text-labeled datasets like CelebV-HQ [Zhu, 2022] and CelebV-Text [Yu, 2023] have facilitated the development of text-conditioned generative models, such as Media2Face [Zhao, 2024]. 

%However, the nuances styles of a facial motion sequence during speech are often difficult to be transcribed to text. As a result, protocols for generating textual labels often default back to categorical emotion labels [Yu, 2023][Zhao, 2024], limiting the ability to author more nuanced or expressive outputs.

\subsection{Example-based Generative Model}
Example-based conditioning is an alternative way to specify the desired style of a generative model. It is commonly used in text-to-speech, since an example audio can intuitively specify a rich set of features at once \cite{zaidi_daft-exprt_2022, zhang_learning_2019, wang_style_2018}. This approach has also shown promise in co-speech gesture generation \cite{ghorbani_zeroeggs_2022} and music-conditioned dance generation \cite{valle-perez_transflower_2021}. 
% \epfix{"dance style??}

In image-based lip-sync models, an input image can be used not only to establish the appearance, but also the expression and style of the speaker \cite{xu_vasa-1_2024}\cite{zhang_sadtalker_2023}\cite{zhou_makeittalk_2020}. However, their main challenge relies in maintaining the image integrity rather than copying the desired motion characteristics. 
% For 3D lip-sync models, Imitator [Thambiraja, 2022] first trains a motion prior that generate un-stylized lip-sync motion, then for stylized speech, it uses iterative optimization to learn an affine transformation in latent feature space to adapt to unseen styles. However, it can only adapt shapes of visemes, but not upperface nor motion trajectories limiting its use to generating personal visemes. 
% Diffposetalk [Sun, 2024] leverages the similarities between motion from the same video clip and differences across different clips by using contrastive loss to train a motion encoder that translates arbitrary clip of motion into a fixed-size latent feature that best represent each clip. And use conditional diffusion on top of it. Improving result upon imitator. However, contrastive loss does not capture similarity between clips (i.e. if two video both contain angry speech, contrastive loss would encourage style encoder to generate vastly different style embeddings), which also makes more appropriete for learning personal speech styles in datasets consisting long clips of unique speakers. 
For 3D lip-sync models, Imitator 
\cite{thambiraja_imitator_2022} first learns a motion prior, that produces un-stylized lip-sync motion, to handle stylized speech, it then iteratively optimizes an affine transformation in the latent space to adapt to unseen styles. However, this approach can only adapt viseme shapes while neglecting upper-face expression and global motion trajectories, limiting its applicability to personal viseme generation. 
This approach also requires model training for each target style.

DiffPoseTalk \cite{sun_diffposetalk_2023} leverages the inherent similarities within motion from the same video clip and the differences across clips through the use of contrastive loss. Using a contrastive loss, they train a motion encoder that compresses a video clip’s motion into a fixed-size latent feature, which is used as conditioning for a diffusion model. Nonetheless, the contrastive loss cannot fully capture stylistic similarities among different clips. For example, if two clips depict \emph{angry} speech, the model is still encouraged to produce dissimilar embeddings. As a result, DiffPoseTalk is more suited to learning individualized speech styles from datasets containing long clips of unique speakers, rather than arbitrary-length videos containing different emotional deliveries.
%LGH: need some work

\subsection{Latent Diffusion Models}
Diffusion models, originally developed for image generation, have become increasingly popular in motion generation, with applications in both speech-driven facial animation \cite{xu_vasa-1_2024}\cite{zhao_media2face_2024} and body animation systems \cite{karunratanakul_optimizing_2024}\cite{tevet_human_2022}. A key advantage of diffusion models is their ability to model many-to-many relationships, effectively avoiding regression to the mean and enabling the generation of diverse and expressive motions \cite{yang_smgdiff_2024}\cite{tevet_human_2022}.
Latent diffusion models build on this foundation by simplifying the learning process. Instead of operating in the original output space, these models make predictions in a latent space, which reduces complexity and improves efficiency \cite{rombach_high-resolution_2022}. 

For 3D facial motion, latent spaces that disentangle expression-related geometry from identity-related geometry are particularly advantageous, as they allow the model to focus on generating motion without having to simultaneously preserve identity features.
The FLAME 3D morphable model \cite{li_learning_2017} is one of the most widely used disentangled latent spaces for facial motion generation and serves as the basis for models like DiffPoseTalk \cite{sun_diffposetalk_2023}, Emote \cite{danecek_emotional_2023}, and EmoTalk \cite{peng_emotalk_2023}. Beyond FLAME, other approaches such as Media2Face \cite{zhao_media2face_2024} and VASA-1 \cite{xu_vasa-1_2024} have trained their own disentangled feature spaces, demonstrating that the key factor for success lies in disentanglement itself rather than reliance on a specific latent representation.
In this work, we learn latent diffusion on the disentangled latent space from SEREP \cite{josi_serep_2024}, as it captures more expressive facial motion compared to FLAME-based methods like \cite{danecek_emoca_2022}, being more suitable for evaluating the transfer of talking styles.

\section{Design Motivations from User Interviews}
\label{sec: design Motivatoin}
To inform our design process for controllable facial animation generation, we conducted interviews with five professional animators at a major game studio company. Each hour-long interview followed a structured protocol covering current workflows, pain points, and reactions to different control paradigms.

All subjects recognized the need for improved tools for automated facial animation, citing an increasing realism gap between generated animations and modern face capture technologies. While subjects appreciated recent approaches such as \cite{xing_codetalker_2023} for generating more natural facial movements, they noted limitations in control and emotional expressiveness.

Preferences for controllability clustered around two main use cases: lower-tier animations (e.g., gameplay sequences) and mid-tier animations (scripted sequences). For lower-tier animations, fully automated solutions are preferred, with minimal time spent on control (e.g., providing a tag). For mid-tier animations, animators prefer more control and generation at interactive-time, starting with an initial automated generation that they can then refine with fine-grained controls.

Example-based approaches emerged as a suitable solution for both use cases. For lower-tier animations, a set of pre-defined videos can be used as tags. This offers more flexibility than traditional tag-based approaches, as new tags can be defined without model re-training. For mid-tier animations, these approaches provide finer control by allowing animators to select specific emotional nuances or express hard-to-describe styles through examples.

\section{Method}
\label{sec:model} 

Our proposed model takes as input an audio sequence $\boldsymbol{A}$, an optional style reference $\boldsymbol{X}_\text{style} = [x^1, ...  ,x^M]$, and generates a new motion sequence $\hat{\boldsymbol{X}}_0 = [x^1, ...  ,x^N]$ such that it lip-syncs to the audio sequence while adhering to the style from $\boldsymbol{X}_\text{style}$.
To achieve this, we jointly train three key components: (1) a variational autoencoder (VAE) that maps a motion clip into a time-invariant style feature \(\boldsymbol{s}\), (2) a style decoder that generates a set of static poses \(\boldsymbol{B} = [b^1, \dots, b^K]\), which we name \emph{style basis}, 
%which modulate the conditional diffusion process, 
and (3) a conditional latent diffusion model that predicts the primary motion \(\Delta \hat{\boldsymbol{X}}\) alongside a style modulation signal \(\boldsymbol{\alpha}\), which governs the influence of the style basis on the primary motion over time.
We operate in the latent space of SEREP \cite{josi_serep_2024} \ep{which disentangles speaker identity and expression features}. After generating the sequence $\hat{\boldsymbol{X}}_0$, we use the mesh decoder from SEREP to obtain the final mesh animation.
% Note that our proposed model operates in the latent space of SEREP \cite{josi_serep_2024}. Any motion sequence in video space, either ground truth motion or style references can be processed into latent code sequences $\boldsymbol{X_\text{style}}$ using the pre-trained SEREP encoder, and we transform the output latent code $\boldsymbol{X_0}$ of our model into mesh space using pre-trained SEREP decoder.

%\epfix{I think we use "basis" and "bases" in different situations. Do we have more than one basis?}

\begin{figure}[h]
    \centering
    \includegraphics[width=1.0\linewidth]{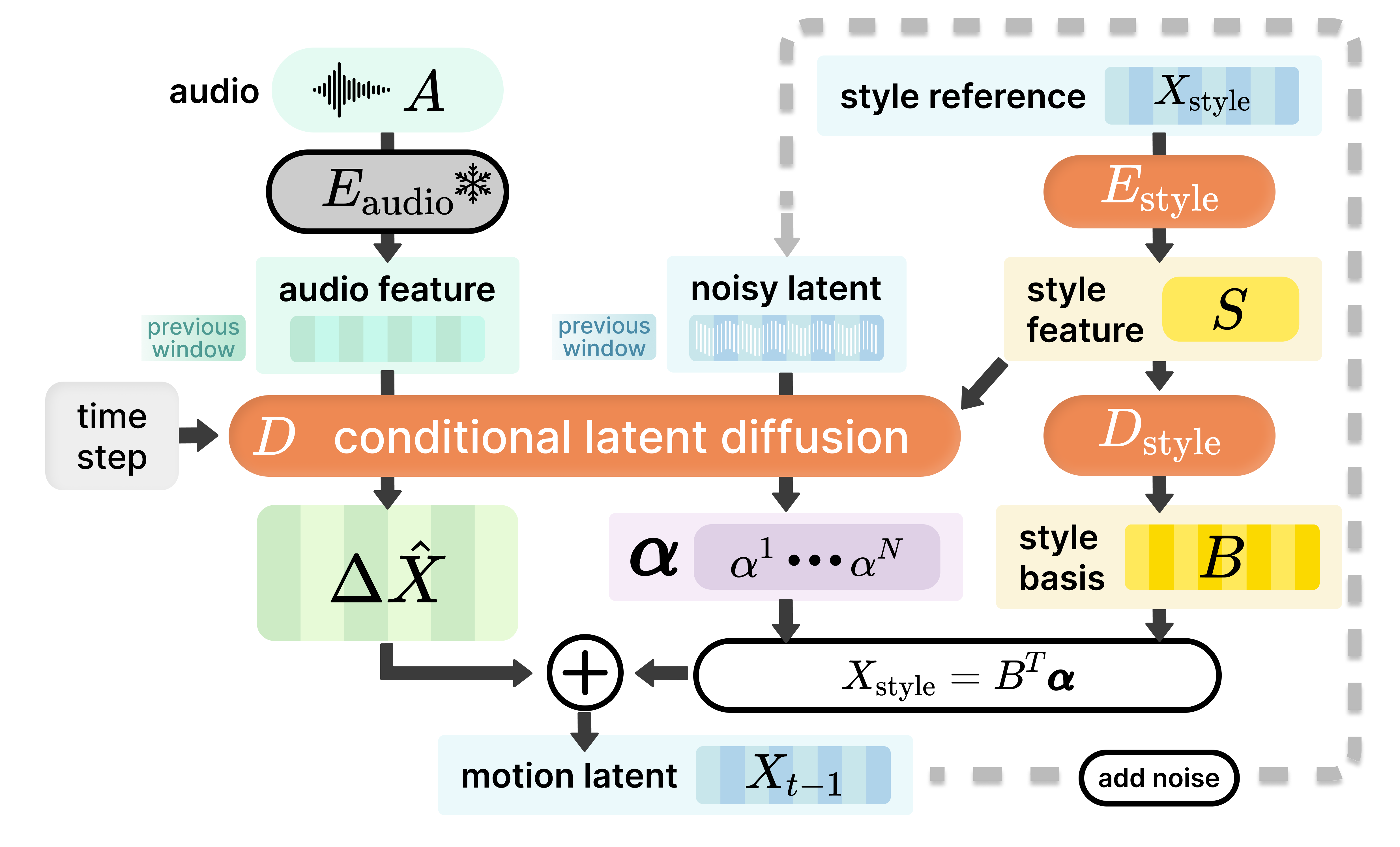}
    \caption{Overview of the proposed method. An encoder extracts style features $S$ from a style reference clip. These are used, in combination with audio features, to condition a latent diffusion model. The diffusion model outputs the primary motion $\Delta \hat{\boldsymbol{X}}$ and a vector \(\boldsymbol{\alpha}\), that modulates the use of the style basis $\boldsymbol{B}$. These are combined to produce the final animation in latent space. This process is repeated for T diffusion steps.}
    % \caption{\ep{A schematic illustration of the interaction between the style encoder, style decoder, and transformer decoder during a single reverse diffusion step.}}
    \label{fig:architecture}
\end{figure}

% Our diffusion model is conditioned on both audio $\boldsymbol{A}$ and a time-invariant style feature $\boldsymbol{s}$ generated by a variational auto-encoder (VAE). During each diffusion step, the model denoise noisy latent $\boldsymbol{X}_n = [x^1, ...  ,x^N]$ to produce $\boldsymbol{X}_{n-1} = [x^1, ...  ,x^N]$ based on our proposed \emph{style basis} representation. During inference, the model can produce motion 
\subsection{Conditional Latent Diffusion Formulation}

We adopt a latent diffusion model inspired by approaches such as VASA-1 \cite{xu_vasa-1_2024} and Media2Face \cite{zhao_media2face_2024}. Rather than directly predicting 3D geometry, the model operates in an expression latent space, which constrains the problem and focuses on compact, meaningful representations. 
% A pre-trained decoder \(D_{\text{latent}}(\cdot)\) is then used to reconstruct the animation sequence from the model’s output, \(\hat{\boldsymbol{X}}_0\). In our experiments, we leveraged the expression space and pretrained decoder of SEREP \cite{josi_serep_2024}.

Diffusion models consist of two Markov chains: a forward diffusion process that progressively adds Gaussian noise to a signal \(\boldsymbol{X}_0\), resulting in pure noise \(\boldsymbol{X}_T\), and a reverse diffusion process \(p(\boldsymbol{X}_{t-1} \mid \boldsymbol{X}_t, t, \boldsymbol{C})\) that reconstructs the clean signal through iterative denoising steps \cite{ho_denoising_2020}. The reverse diffusion process is conditioned on a vector \(\boldsymbol{C}\) and the diffusion time step \(t\). 

To approximate the reverse diffusion process, we train a model \(\mathcal{M}(\boldsymbol{X}_t, t, \boldsymbol{C})\). During inference, this model predicts the motion signal from an arbitrary starting noise \(\boldsymbol{X}_T\). Similar to prior works \cite{xu_vasa-1_2024,sun_diffposetalk_2023,tevet_human_2022}, our model is trained to predict the latent signal and minimize the simple loss $\mathcal{L}_\text{simple}$ over the joint distribution of condition $C$ and clean signal $\boldsymbol{X}_{0}$:
\begin{align}
    \mathbb{E}_{t \sim \mathcal{U}[1, T], \boldsymbol{X}_{0}, \boldsymbol{C} \sim q(\boldsymbol{X}_{0}, \boldsymbol{C})} \left[ \|\boldsymbol{X}_{0} - \mathcal{M}(\boldsymbol{X}_t, t, \boldsymbol{C})\|^2 \right].
\end{align}
To accommodate speech of arbitrary length, we use a windowing strategy  \cite{xu_vasa-1_2024, sun_diffposetalk_2023}, where the input audio is segmented into fixed-length windows of size $N_w$. We perform diffusion on each window sequentially. To ensure smooth transitions between windows, we include \(N_p\) frames of audio and motion features from previous windows as part of the conditioning signal. We further regularize the motion with velocity loss $\mathcal{L}_\text{vel}$ \cite{cudeiro_capture_2019} and smoothness loss $\mathcal{L}_\text{smooth}$ to ensure that the model is learning the correct dynamic and prevents jerky motion:
\begin{align}
&\mathcal{L}_\text{vel} = \|
({\boldsymbol{X}}_{N_p+1:N_w}-{\boldsymbol{X}}_{N_p:N_w-1}) - (\hat{\boldsymbol{X}}_{N_{p}+1:N_w}-\hat{\boldsymbol{X}}_{N_{p}:N_{w}-1)})
\|^{2},
\\
&\mathcal{L}_\text{smooth} = \|
\hat{\boldsymbol{X}}_{N_{p}+2:N_w} - 
\hat{\boldsymbol{X}}_{N_{p}+1:N_w-1} +
\hat{\boldsymbol{X}}_{N_{p}:N_w-2}
\|^2
\end{align}

\subsection{Conditioning Signal}
Our conditioning signal combines multiple components: an audio feature \(\boldsymbol{A}\), style features \(\textbf{s}\), the last $N_p$ frames of motion latents \(\boldsymbol{X}_{\text{prev}}\) from the previous window, and their corresponding audio features \(\boldsymbol{A}_{\text{prev}}\). The audio features are produced using a pre-trained HuBERT audio encoder \cite{hsu_hubert_2021}, with all layers kept frozen to avoid over-fitting to the training data. Meanwhile, the style feature \(\textbf{s}\) is extracted from an arbitrary-length style reference by the variational auto-encoder \(E_{style}\), which uses a transformer encoder to first generate per-frame style features \ep{that considers the varying importance of each frame}, then mean-pooled across time steps to ensure it can handle clips of arbitrary length. Within the VAE framework, \(E_{style}\) predicts the mean \(\boldsymbol{\mu}\) and variance \(\boldsymbol{\sigma}^2\) of the latent distribution. The style vector \(\textbf{s}\) is then generated using the re-parameterization trick:
\[
\textbf{s} = \boldsymbol{\mu} + \boldsymbol{\sigma} \odot \boldsymbol{\epsilon}, \quad \boldsymbol{\epsilon} \sim \mathcal{N}(0, \mathbf{I})
\]
This approach allows for differentiable sampling during training. To regularize the latent space, the encoder is optimized with the Kullback-Leibler divergence loss:
\[
L_{KL} = \frac{1}{2} \sum_{i=1}^{d} \left( \mu_i^2 + \sigma_i^2 - \log \sigma_i^2 - 1 \right)
\]
This loss ensures that the learned distribution remains close to a standard normal prior, promoting a smooth and continuous style space. 
Further implementation details can be found in the supplementary code. 

\subsection{Guiding Diffusion with a Style Basis}
\label{sec: style basis}
% In prior works, the term "style" has referred to both the speech style when expressing different emotions [Peng, 2023] [Daněček, 2023] and the personal speech style of individuals [Thambiraja, 2022][Sun, 2024]. We don't make such distinction, in our work, we simply define style as time-invariant property of a given clip of speech performance. 
% When we first started training with an end-to-end style encoder, with our probabilistic crisscross method, we found that it often leads to the style encoder compressing the entire motion sequence rather than picking out the time-invariant feature across the clip, which leads to incoherent lip-sync during inference. We proposes style bases to counter this problem of leakage. By enforcing the style code to learn a time-invariant aspect of the speech, it trains a more coherent and robust feature space \ep{maybe show t-SNE plot here}. 

% We conducted preliminary experiments with a VAE-based style encoder trained in an end-to-end fashion and observed that it often led to the style encoder $E_{\text{style}}$ compressing the entire style sequence $\boldsymbol{X}_{\text{style}}$ into the style code. This resulted in the model learning sequence-specific features rather than the time-invariant "delivery style" of the clip. 

We define \emph{style basis} as:
$\boldsymbol{B} = [\boldsymbol{b}_1, \boldsymbol{b}_2, \ldots, \boldsymbol{b}_K]$, 
where each \(\boldsymbol{b}_k \in \mathbb{R}^{1 \times D}\) is a latent expression code. Conceptually, these basis vectors capture low-frequency or persistent expressions that are characteristic of the speech mannerism from the style reference. \ep{This is inspired by how animators in gaming industry overlay co-speech expressions on top of high frequency lip sync}. These are generated from the style latent using the style decoder $D_\text{style}$, 
%where each \(\boldsymbol{b}_k\) is a static expression vector in \(\mathbb{R}^{1 \times D}\). Conceptually, these basis vectors capture low-frequency or persistent expressions that are characteristic of the speech mannerism from the style reference. These are generated from the style latent using the style decoder $D_\text{style}$, 

\begin{figure}[h]
    \centering
\includegraphics[width=1\linewidth]{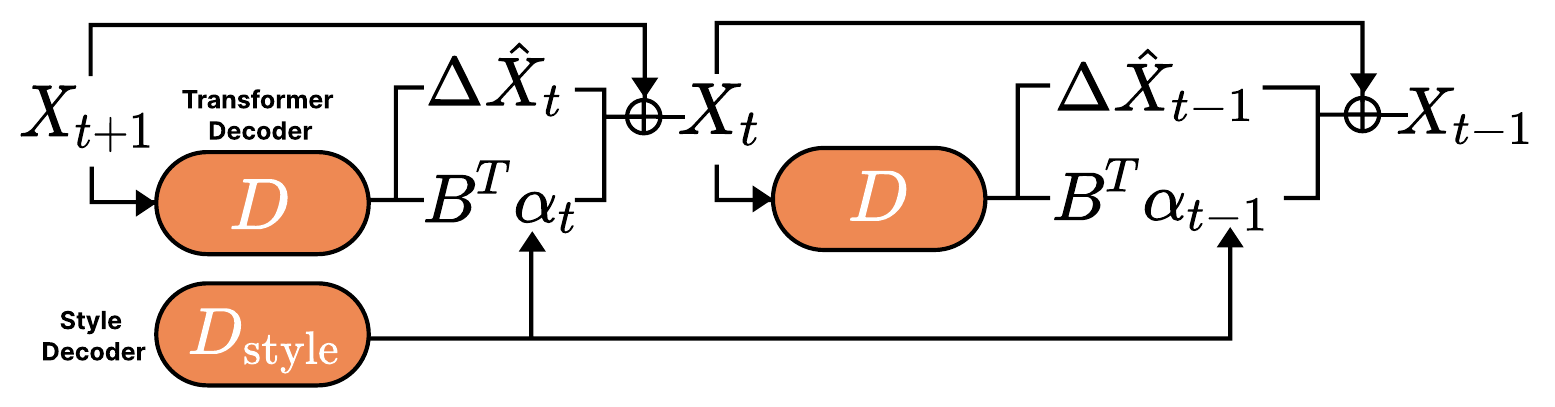}
    \caption{The style basis $\boldsymbol{B}$ is iteratively incorporated into primary motion throughout the reverse diffusion process, modulated by $\alpha_t$. Note the reverse diffusion process works in decreasing time step from T to 0. }
    % \caption{\ep{A schematic demonstrating how style basis $\boldsymbol{B}$ is iteratively incorporated into primary motion throughout the reverse diffusion process, modulated by $\alpha_t$. Note the reverse diffusion process works in decreasing time step from T to 0.} }
    \label{fig: recursion figure}
\end{figure}

As shown in Figure \ref{fig:architecture}, we model the diffusion output \ep{\({\boldsymbol{X}} \in \mathbb{R}^{N \times D}\)} 
as the sum of (1) a primary motion \ep{\(\Delta\hat{\boldsymbol{X}} \in \mathbb{R}^{N \times D}\)} and 
(2) a linear combination of style basis vectors scaled by a time-varying style modulation signal
\ep{\(\boldsymbol{\alpha}_i \in \mathbb{R}^{N \times 1}\)}:
\begin{align}
    \hat{\boldsymbol{X}} 
    = \Delta\hat{\boldsymbol{X}} 
    + \sum_{i=0}^{K} \boldsymbol{\alpha}_{i} \,\boldsymbol{b}_i.
\end{align}
Each coefficient \(\boldsymbol{\alpha}_i\) can vary over time, modulating how much the style basis \(\boldsymbol{b}_i\) influences the primary motion throughout the animation. Through the iterative nature of diffusion, the influence of the style basis is incorporated into the primary motion, as shown in Figure \ref{fig: recursion figure}.

% Practically, we found that while $\Delta \hat{\boldsymbol{X}}$ is focused on captures the lip-sync motion, $\alpha$ and the static basis animates the upper face and alters the lip shape, as shown in the supplementary video \textcolor{red}{(XX:XX-XX:XX)}. } In our model, the style bases $\boldsymbol{B}$ are generated by a 3-layered feed forward network $D_{style}$ from on the output of the style encoder (see Figure \ref{fig:architecture}).  

\subsection{Training Strategy}
The training objective for our model is composed of the previously specified simple loss and various regularization objectives.
\begin{align}
\mathcal{L}=\lambda_\text{simple}\mathcal{L}_\text{simple}+\lambda_\text{vel}\mathcal{L}_\text{vel}+\lambda_\text{smooth}\mathcal{L}_\text{smooth}+\lambda_\text{KL}\mathcal{L}_\text{KL}
\end{align}
To train our window-based model, the network must address two scenarios: (a) generating the initial window conditioned on learnable start features, and (b) generating subsequent windows conditioned on speech features and motion parameters from the preceding window. To handle both, we train the model using pairs of examples. Specifically, we sample an audio clip and a motion clip of total length $2N_w$ (zero-padded if shorter) and split them into two windows of length $N_w$, with the two windows covering scenarios (a) and (b), respectively. 
In \cite{sun_diffposetalk_2023}, it is assumed that the style is consistent across the two windows. Their method employs cross-window conditioning, where window $b$'s motion serves as the style clip to reconstruct window $a$'s motion using window $a$'s audio. However, emotionally expressive speech often involves varying delivery intensities throughout the clip, making this assumption unreliable. To address this, we alternate between cross-conditioning, which uses motion features from the opposite window as the style input, and self-conditioning, which uses motion features from the same window. 
This strategy mitigates the risk of style leakage which can happen with pure self-conditioning, where the style encoder may encode sequence-specific motion features instead of capturing time-invariant delivery speech styles. 
% It also improves upon pure cross-conditioning, empirically enhancing generalization across diverse motion styles and scenarios. 
% We find that the crisscross arrangement prevents the model from merely compressing and decoding the same window, while incorporating the non-crisscrossed style clip encourages the model to capture and utilize specific expressions or iconic frames present in the style clip. 

Similar to \cite{xu_vasa-1_2024}, to allow for more nuanced control, and the option to generate un-stylized speech, we also incorporated \emph{classifier free guidance (CFG)} \cite{ho_classifier-free_2022}. During training, we randomly drop $10\%$ audio and style guidance by replacing frames of the corresponding conditioning signal with a learned null condition $\varnothing$. During inference, we apply:

\begin{equation}
    \hat{\boldsymbol{X}}_0 = (1 + \sum_{c\in \boldsymbol{C}} \lambda_c)\cdot \mathcal{M}(\boldsymbol{X}_t, t, \boldsymbol{C}) - \sum_{c\in \boldsymbol{C}} \lambda_c \cdot \mathcal{M}(\boldsymbol{X}_t, t, \boldsymbol{C}|_{c=\varnothing})
\end{equation}

Where $\lambda_c$ is the cfg scale for the corresponding condition, and $\boldsymbol{C}|_{c=\varnothing}$ indicates that the corresponding condition is replaced with null condition $\varnothing$. 

\section{Experimental Protocol}
In this section, we detail our experimental protocol, intending to evaluate the impact of different methods for style transfer. For this purpose, we curated a dataset of expressive and stylized speech. To isolate the impact of style transfer methods, we use the same dataset, facial expression representation, overall model architecture and training protocol for all experiments, varying only how style is computed from the style reference, and how it is used for generation.

\paragraph{\textbf{Dataset.}}
We conducted our experiments using a combination of the CelebV-Text   \cite{yu_celebv-text_2023} and the RAVDESS datasets  \cite{livingstone_ryerson_2018}. Both datasets primarily consist of videos featuring a single speaker addressing the camera. The RAVDESS dataset contains videos of the same speakers expressing different emotions, while CelebV-Text consists of unique speakers in each video. Videos in CelebV-Text are typically between 10 to 15 seconds long, whereas RAVDESS videos range from 3 to 6 seconds.

CelebV-Text also includes videos with non-speaking individuals, masked speakers, speakers eating, and speakers facing away from the camera. To filter out these instances, we utilized the dataset's textual tags (e.g., “masked,” “eating”) and queried only for videos that involved speech, singing, whispering, or reading. Additionally, we used head pose and facial bounding box detection from MediaPipe \cite{lugaresi_mediapipe_2019}  to exclude videos where the speaker's face was angled more than 45 degrees from the camera direction or contained significant head pose jumps. From the remaining videos, we randomly sampled 26.5 hours of content to match the data scale used in prior work, DiffPoseTalk \cite{sun_diffposetalk_2023}, which we use for comparison. The entire RAVDESS dataset (3 hours) was used as it does not suffer from video quality issues.

To extract expression coefficients, we first cropped video frames using MediaPipe facial bounding boxes, then applied SEREP \cite{josi_serep_2024} to compute the facial expression, represented as 64-dimension expression codes, while MediaPipe is used to extract 3-dimension head poses. The full latent representation $X$ at every frame therefore contains 67 dimensions.

For the train-validation split, we randomly selected $90\%$ of the data for training and $10\%$ for validation. 
% For RAVDESS, we used a leave-two-out split strategy, training only on the emotions "neutral," "calm," "happy," "angry," "fearful," and "disgust," while reserving "sad" and "surprised" for validation \epfix{did we use this? if not we should drop this text}. We’ve additionally collected and processed \epfix{10} iconic performances from well-known movies, interviews and TV shows to be used to qualitatively showcase our model.

\paragraph{\textbf{Implementation details.}}
Our model architecture is inspired by DiffPoseTalk \cite{sun_diffposetalk_2023}. The style encoder features a 4-layer transformer encoder (hidden size $d=512$) with 4 attention heads, followed by 2 fully connected layers with 512 and 256 output features each.
The final output is the mean and standard deviation of the style code, each of dimension $128$, that are used in the re-parameterization trick for the VAE.
%The final output is a style feature of size $D=256$, where the first 128 features represent the mean ($\mu$) and the remaining 128 features represent the standard deviation ($\sigma$), which are used in the re-parameterization trick for VAE. 
For the audio encoder, we utilize the HuBERT model \cite{hsu_hubert_2021} with its feature extractor frozen, followed by a linear layer mapping the output to 512 features.

The motion decoder consists of an 8-layer transformer decoder (with a hidden dimension of $512$) with 8 cross-attention heads, attending to both the noisy feature and the audio. The style features are appended to the beginning of the noisy features, allowing the decoder to attend to them at every step. For the style decoder, we use $K$ sets of 3 fully connected layers to map the style features into $K$ style bases, with $K=4$ for our evaluation experiments. We use a classifier-free guidance strength of 1 ($\lambda_c=1$) to generate all examples. 

We train all modules end-to-end using the Adam optimizer with a learning rate of 0.0001 and a batch size of 16. Data samples are prepared using a window size of $N=100$ frames with an overlap of $N_p=10$ frames.
The loss function weights are set as follows: $\lambda_{\text{vel}} = 0.5$, $\lambda_{\text{smooth}} = 5 \times 10^{-6}$, and $\lambda_{\text{KL}} = 1 \times 10^{-6}$, \ep{experimentally selected to prevent mode collapse during training and kept constant for all experiments}. The model is trained on a single NVIDIA A4000 GPU for 450,000 iterations (selected with an early stop mechanism), which takes approximately 4 days.

% \lgh{All trained models predict sequences of both the 64-dimension expression code and 3 parameters for head pose. As in previous work, the quantitative metrics for lip-sync and upper-face movement ignore head motion. We also ignore head motion in the user study to ensure that participants and evaluators could focus solely on facial expressions and lip synchronization.}

%All models were trained to predict head motion using the same approach shown in DiffPoseTalk \cite{sun_diffposetalk_2023}. However, since our primary focus is on facial performance, for consistency in qualitative analysis and user studies, we zeroed out the head motion to ensure that participants and evaluators could focus solely on facial expressions without distraction. We use a cfg strength of 1 ($\lambda_c=1$) to generate all of our examples.  

\paragraph{\textbf{Baselines}}
%We compare our model against the state-of-the-art style-guided lip-sync model DiffPoseTalk , as well as ablated versions of our proposed model. The ablations include an style agnostic model that ignores both the style encoder and style decoder (unconditioned), a pure VAE model trained with pure cross-condition (no style basis), and lastly the proposed model trained with pure cross-condition (no mix condition). All models are trained on the proposed dataset for up to 500,000 iterations, with the best-performing models selected using an early stopping mechanism.

We compare our model against two baselines: the state-of-the-art DiffPoseTalk model \cite{sun_diffposetalk_2023}, and a baseline we refer to as VAE-baseline. 
DiffPoseTalk uses a style-encoder that is pre-trained using a contrastive loss. It then trains a diffusion model conditioned on the style obtained from this pre-trained encoder, which remains frozen throughout the training process. For VAE-baseline, we implemented a VAE style encoder that is jointly trained with the generative model, in a similar vein to \cite{ghorbani_zeroeggs_2022, zaidi_daft-exprt_2022}. We employ the same diffusion architecture as the other models for fair comparison, \ep{utilizing Denoising Diffusion Probabilistic Models (DDPM) for inference \cite{ho_denoising_2020}, with 1000 denoising steps.} Our model differs from VAE-baseline by the incorporation of the style basis, and how we employ a mix of cross-window and self-window conditioning.
\ep{For all evaluations, style references are utilized by all models being compared, ensuring that performance gains are not attributable to any model having access to additional information.}
% It is important to note that existing models for style-guided lip-sync generation are limited, and DiffPoseTalk is the best state-of-the-art model in this domain. This highlights the novelty of our work and the need for further development in style-guided generation approaches.

\section{Results}
Models are evaluated quantitatively using established metrics, qualitatively with examples, and perceptually with a set of user studies. 

\subsection{Quantitative Results}
To quantitatively evaluate the models, we use four established metrics: vertex mean squared error (MSE), lip vertex error (LVE) \cite{richard_meshtalk_2022}, upper-face dynamics deviation (FDD) \cite{xing_codetalker_2023}, and mouth opening difference (MOD) \cite{sun_diffposetalk_2023}. MSE and LVE measure the per-vertex $L_2$ error between the target and generated motions, with MSE focusing on the entire face and LVE specifically targeting the vertices around the lips. FDD computes the difference in the standard deviation of each upper-face vertex, serving as a proxy for the dynamics of upper-face motion. MOD evaluates the difference in the extent of the mouth openings, which is indicative of the speech style \ep{(e.g. shout vs whisper)}. 

\ep{For each test example, we source the style reference from an adjacent but non-overlapping window from the same video as the ground truth. As the clips are adjacent, the style reference contains similar style as the ground truth, ensuring the model has access to the correct style information. 
By selecting the style clip to be non-overlapping with ground truth, we ensure that we are evaluating style-conditioned generation capabilities rather than motion recreation.}

The quantitative results are presented in Table \ref{tab:metrics_comparison}. Our method demonstrates improved performance across all metrics compared to baselines, indicating its ability to generate stylistically accurate motion in both the upper and lower face.

The ablation study further shows that most of the performance in LVE and and MSE can be attributed to the inclusion of the style basis, while the inclusion of cross condition allows the model to better capture the dynamics of the upper face, improving FDD. 
% \begin{table}[ht]
% \vskip -2mm
% \centering
% \caption{Quantitative comparison of the proposed model and baselines. Lower values indicate better performance across all metrics. \textbf{Bolded} values represent statistically significant improvements over all other models (\( p < 0.05 \)).}
% \vskip -2mm
% \label{tab:metrics_comparison}
% \begin{tabular}{lcccc}
% \toprule
%  & MSE & LVE & FDD & MOD \\
% Model & (mm) $\downarrow$ & (mm) $\downarrow$ & ($\times 10^{-5}$m) $\downarrow$ & (mm) $\downarrow$ \\
% \midrule
% Unconditioned &
%     $4.5$ & 
%     $11.0$ & 
%     $52.4$ & 
%     $3.2$    
% \\
% DiffPoseTalk &
%     $3.7$ &  
%     $9.7$ &  
%     $44.8$ & 
%     $3.4$    
% \\
% VAE-baseline &
%     $3.6$ &  
%     $9.1$ &  
%     $44.6$ & 
%     $3.1$   
% \\
% \textbf{MSMD (ours)} &
%     $\textbf{3.3}$ & 
%     $\textbf{8.3}$ & 
%     $\textbf{36.5}$ & 
%     $\textbf{2.7}$   
% \\
% \bottomrule
% \end{tabular}
% \vskip -2mm
% \end{table}

\begin{table}[ht]
% \vskip -2mm
\centering
\caption{Quantitative comparison of the proposed model and baselines. Lower values indicate better performance across all metrics. \textbf{Bolded} values represent statistically significant gain over all other models (\( p < 0.05 \)).}
% \vskip -3mm
\label{tab:metrics_comparison}
\begin{tabular}{lcccc}
\toprule
 & MSE & LVE & FDD & MOD \\
Model & (mm) $\downarrow$ & (mm) $\downarrow$ & ($\times 10^{-5}$m) $\downarrow$ & (mm) $\downarrow$ \\
\midrule
Unconditioned &
   $4.50$ &  % 0.0043701 * 1000
   $11.0$ & % 0.0107068 * 1000
   $52.4$ &  % 0.0000059 * 10^5
   $3.24$   % 0.0031468 * 1000
\\
DiffPoseTalk &
   $3.86$ &  % 0.0037985 * 1000
   $10.0$ &  % 0.0098586 * 1000
   $48.3$ &  % 0.0000047 * 10^5
   $3.46$   % 0.0033892 * 1000
\\
VAE-baseline &
   $3.61$ &  % 0.0035310 * 1000
   $9.13$ &  % 0.0088995 * 1000
   $47.4$ &  % 0.0000046 * 10^5
   $3.10$   % 0.0030000 * 1000
\\
\textbf{MSMD (ours)} &
   $\textbf{3.46}$ &
   $\textbf{8.54}$ &
   $\textbf{43.4}$ & 
   $\textbf{2.96}$
\\
\bottomrule
\end{tabular}
\end{table}

\begin{table}[ht]
\centering
\caption{Quantitative comparison of the proposed model and ablations. Lower values indicate better performance across all metrics. \textbf{Bolded} values represent statistically significant gain over all other models (\( p < 0.05 \)).}
% \vskip -2mm
\label{tab:ablation_comparison}
\begin{tabular}{lcccc}
\toprule
Model & MSE & LVE & FDD & MOD \\
      & (mm) $\downarrow$ & (mm) $\downarrow$ & ($\times 10^{-5}$m) $\downarrow$ & (mm) $\downarrow$ \\
\midrule
no style basis &
   $3.61$ &  % 0.0035310 * 1000
   $9.13$ &  % 0.0088995 * 1000
   $47.4$ &  % 0.0000046 * 10^5
   $3.10$   % 0.0030000 * 1000

\\
no cross condition &
    $3.59$ &  
    $8.96$ &  
    $46.3$ & 
    $3.01$    
\\
MSMD (ours) &
   $\textbf{3.46}$ &
   $\textbf{8.54}$ &
   $\textbf{43.4}$ & 
   $2.96$
\\
% note this is running the trained model without CFG rather than training a model without CFG. The model trained with CFG is happening now i will fill it in later.
\bottomrule
\end{tabular}
\end{table}

\subsection{Qualitative Results}
\begin{figure*}[p!]
    \centering
    \includegraphics[width=0.8\linewidth]{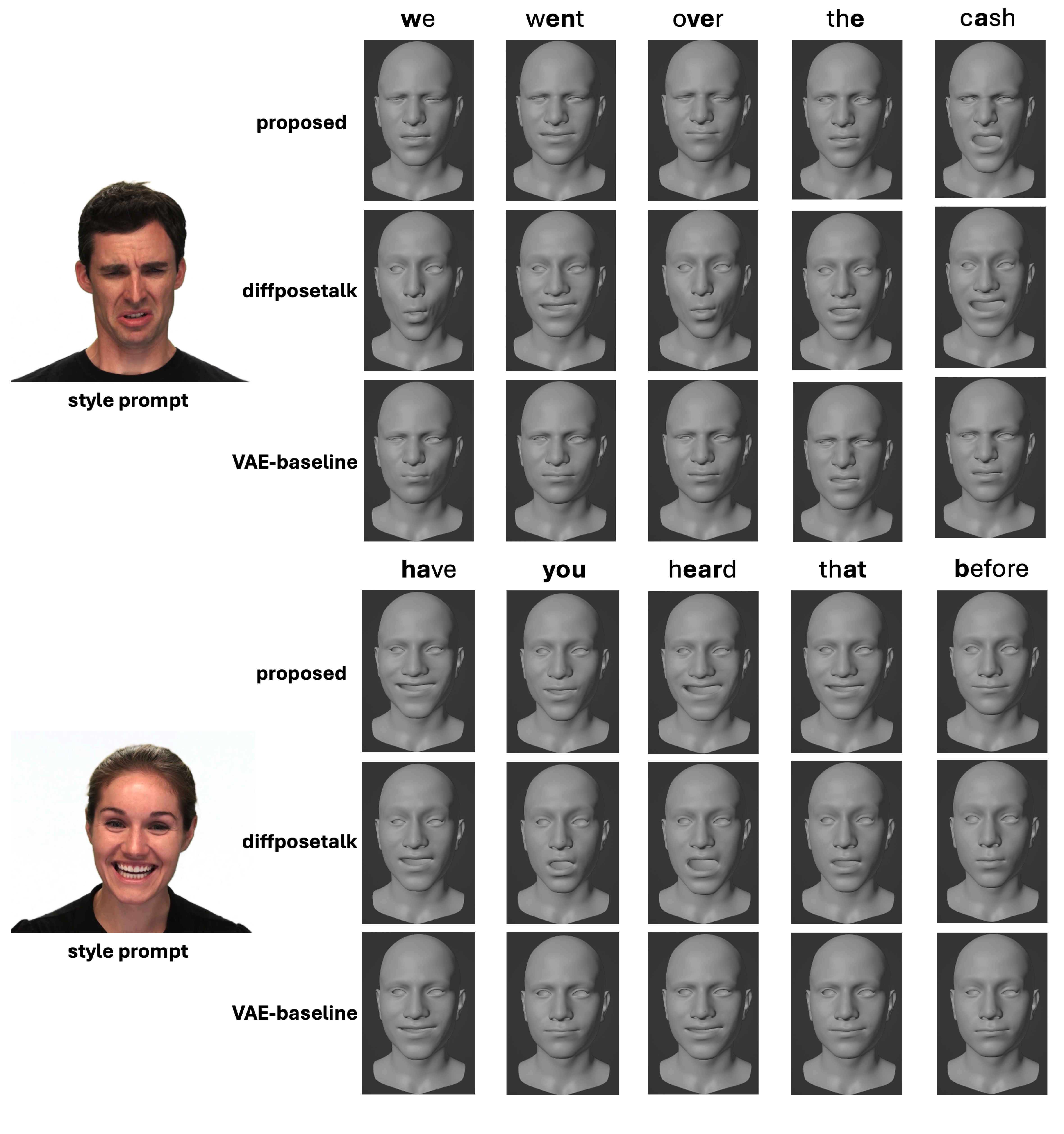}
    \caption{Qualitative results of baselines and proposed models on different style prompts and audio.}
    \label{fig:qualitative_figure}
\end{figure*}

% \begin{figure}
%     \centering
%     \includegraphics[width=\linewidth]{figures/qualitative_comparison.png}
%     \caption{Qualitative comparison of style transfer with baselines}
%     \label{fig:qualitative}
% \end{figure}
 We refer the reader to the supplementary video (at time 3:44) for video results. Figure \ref{fig:qualitative_figure} compares generated frames from our method and the baselines. 
% Video 3:44-4:54.
These examples are generated using emotionally expressive videos from RAVDESS as the style reference,  and audio tracks from CelebV-Text. Notably, this combination lies entirely outside the training distribution, further demonstrating the generalization capability of our method.

\begin{figure*}[p!]
    \centering
    \includegraphics[width=1\linewidth]{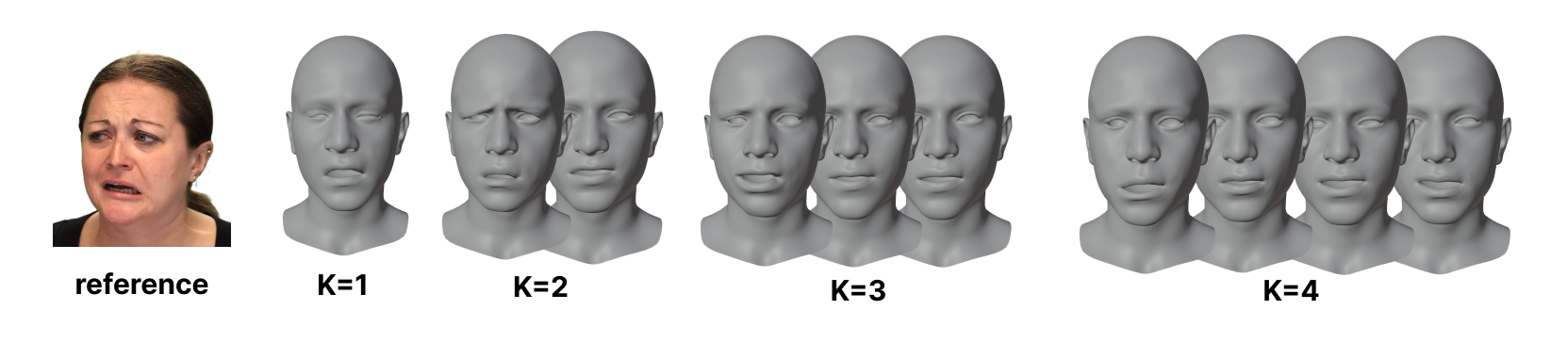}
    \caption{Example of style basis from models of basis size 1, 2, 3, and 4 and the corresponding reference. }
    \label{fig:style_basis}
\end{figure*}

\begin{figure*}[h!]
    \centering    \includegraphics[width=1\linewidth]{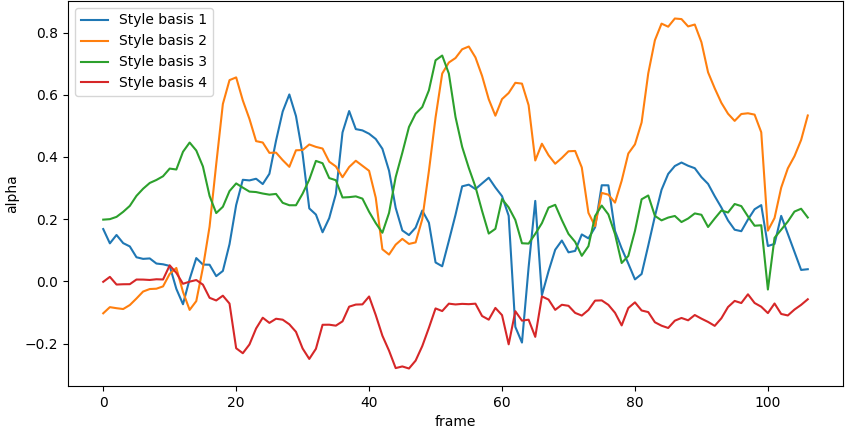}
    \caption{Example of alpha trajectory over time for 4 style basis. }
    \label{fig:alpha_trajectory}
\end{figure*}

In most cases, DiffPoseTalk either introduces exaggerated expressions or fails to accurately capture the intended style references. This limitation is likely due to its reliance on training the style encoder solely with contrastive loss \cite{sun_diffposetalk_2023}, which struggles to capture shared attributes across clips, such as emotions. Similarly, the VAE-baseline model, which excludes the style basis and cross-condition components, produces expressions that roughly align with the reference but frequently omits nuanced upper-face expressions.
In contrast, our proposed model effectively captures the reference style, closely matching both the lower-face dynamics and upper-face expressions. We provide additional results of our model rendered with head motion derived from in-the-wild performances, using a 4-second clip as the style reference. See the supplementary video at time 5:15. \ep{We further provide visualizations of style basis and example trajectory of alpha (Figure \ref{fig:style_basis}, Figure \ref{fig:alpha_trajectory} to provide insights to style basis.}
%Video 0:00-0:16, 5:15-7:00.
% This demonstrates the importance of incorporating the style basis and cross condition mechanisms in achieving more expressive and stylistically accurate results.

\subsection{User Study}
% We conducted user studies to compare the models in terms of style transfer performance, lip-sync accuracy, and the naturalness of motion. 

% \lgh{We carried out two evaluations: One using Ravdess, and one using in-the-wild videos. This evaluates how the models compare on highly emotional speech (from a basic set of emotions in ravdess) and how it is capable of transferring arbitrary styles, that may be hard to describe in words. } 

% \lgh{For each set, we select 10 pairs of reference style clips and audio. When evaluating style, we mute the videos, to ensure that subjects do not pay attention to lip synchronization. Similarly, for evaluating lip-sync and naturalness, we not show the style reference, so that it does not influence their evaluation.}

We conducted user studies to compare the ability of different models on generating animations maintaining the \emph{style} of a style reference, and producing accurate \emph{lip-synchronization}. 

%We chose a total of 10 videos, considering two scenarios: (a) using audio and style references of matching styles; (b) using a reference style clip of a different style than the audio.
We selected 10 videos, considering two scenarios: (a) audio paired with matching style references; (b) audio paired with contrasting style references.
For the first scenario, we consider 5 in-the-wild videos, split the videos in two segments, using the start as style reference, and the remaining as the audio input. The second scenario stress-tests the models on the case where the style reference is vastly different than the audio. In this case, we use 5 highly-emotional videos from RAVDESS as style reference, and select in-the-wild audio that do not match the selected emotion for the clips.

To evaluate style transfer and lip synchronization independently, we designed two separate user studies. For each study, participants were shown the same set of generated animations, rendered without head motion to allow them to focus on facial expressions.

\paragraph{\textbf{Style transfer evaluation}}

In the first study, participants evaluated the system's ability to transfer speaking styles from a source video to the generated animation.
Participants viewed the style reference video at the top of the screen, followed by four different characters (labeled A, B, C, D). Besides showing the animations from the 3 models under test, we selected a random animation from the validation set (with a different style) as a negative anchor. To focus solely on style transfer quality, videos were presented without audio. Participants rated each character on a scale of 0-5 based on how closely it matched these properties of the style reference video: Facial expressions, Speaking style, and Distinctive facial movements.

\paragraph{\textbf{Lip synchronization assessment}}

The lip synchronization study evaluated the temporal alignment between speech and mouth movements. Participants viewed four characters speaking with audio enabled and rated the lip-sync quality on a 0-5 scale, where 0 represented completely mismatched movements and 5 represented perfect synchronization. Besides the 3 models under test, we selected a random animation from the validation set (with different audio) as a negative anchor. To isolate lip-sync quality, participants were instructed to disregard variations in speaking style between characters and focus only on the audio-visual alignment.

Each study was designed to take 10 to 15 minutes. Videos were presented in a looping format to allow participants sufficient time for observation. Participants were encouraged to submit their ratings only when confident in their assessment. This approach allowed for careful evaluation while maintaining reasonable time constraints for the complete study session.  Please refer to the supplementary material for the detailed instructions given to participants.

%\ep{Users are asked to judge each output in terms of style accuracy, and lip-sync accuracy. We define style accuracy as how similar the style of the output animation is to a corresponding reference video, irrespective to speech. While lip-sync accuracy refer to how plausible the generated lip motion is with respect to the speech.} 

% Through preliminary pilots, we found that the tasks of identifying style accuracy and lip-sync accuracy are both very demanding and requires a participant's full focus. For this reason, we split the study into two parts to collect style accuracy and lip-sync accuracy in separate experiments.
% To collect response of style accuracy, we show users the style reference video along with the output animations from each baseline with no sound. We instruct the user to focus only on evaluating style of the delivery and pay no attention to lip-sync.
%For lip-sync accuracy, we display only the model outputs, along with corresponding audio. And we instruct the user to pay attention only to lip-sync while ignoring style of the delivery. For both sets of experiments, the users were asked to rate each performance (0-5). We also employ attention checking and normalization by incorporating a carefully chosen negative example with non-matching lip-sync and style to help calibrate participant responses, providing a clear baseline for identifying poor performance.

\begin{figure}
    \centering
    \includegraphics[width=1\linewidth]{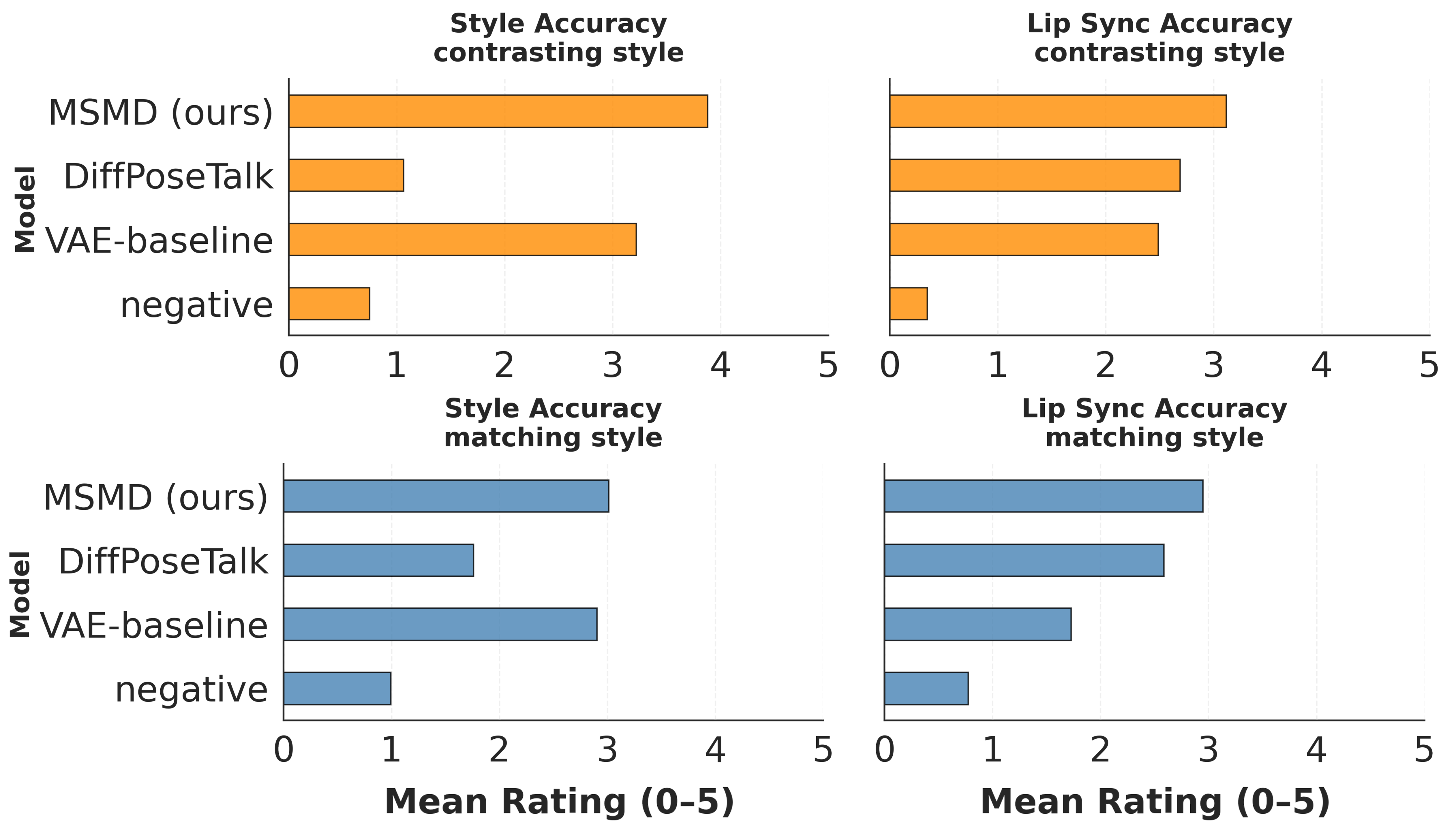}
    \caption{
        User-scored performance comparison of the proposed model and baselines for style transfer and lip sync. Results are shown for contrasting style (top) and matching style (bottom), with mean ratings on a 0-5 scale.   
    % Average score of lip sync accuracy and style accuracy of key baseline models and the negative example, separately by conditions of whether the style reference matches audio delivery style. 
    }
    \label{fig:user-study-result}
\end{figure}
We collected 34 and 32 responses for the style accuracy and lip sync accuracy experiments, respectively. The results of this evaluation are summarized in Figure \ref{fig:user-study-result}. We observe that our model has superior performance both at capturing the style and generating the correct lip sync, especially for capturing highly expressive speech styles from the RAVDESS examples (contrasting style). A set of paired t-tests confirms that our model’s mean ratings are significantly higher than all competing methods ($p<0.0005$).

On the other hand, DiffPoseTalk was unable to capture the speech styles in most cases, showing the limitations of training the style encoder as a separate task with contrastive loss. Both DiffPoseTalk and the VAE-baseline show worse lip sync capability than the proposed model. We hypothesize that our performance gain is due to the style basis representation, which effectively decouples speech from style, preventing style from disrupting essential viseme shapes.

\section{Discussions}

\subsection{Use Case of Style-conditioned Generation}
Style-conditioned generation enables two distinct applications:
First, it can seamlessly integrate into existing pipelines that rely on tag-based animations (see Section \ref{sec: design Motivatoin}). By curating a set of style references to represent various tags, offering greater flexibility. The curated examples can be easily updated with more suitable ones, or the tag library can be expanded to accommodate new styles.

Secondly, style-conditioned generation can be applied to produce dubbed animations while preserving certain aspects of the original performance. For instance, the original performance can serve as the style reference, while using an input audio track in a different language. This approach is more versatile than simply replacing lip movements, as it allows for dubbing over the original performance with audio of varying lengths, maintaining expressive alignment.
\subsection{Limitations}

While our model can mimic the speech style of arbitrary style references to any audio while maintaining lip sync coherence, it does not consider whether the delivery style of the speech, such as tone, pace, and duration matches the delivery style in the reference. For example, the model can map an angry speech style to an audio with an uplifting tone which would generate uncanny animation (see the supplementary video at time 4:54). 

Furthermore, our model is designed based on the definition that delivery style is something invariant throughout the delivery of a line of speech. In the case where speech transitions from one style to another (such as from a neutral expression to an angry expression), we don't have a way to innately control the transition, and the animator would have to manually break the audio into windows and use different style for each window.

Lastly, our model does not consider replicating isolated instances of iconic gestures such as the eyebrow raise of Dwayne "the rock" Johnson, or eye-rolling. Which can be considered very significant to the style of a performance. A potential future direction could be finding a way to incorporate these facial gestures into generation. 
\subsection{Future Works}

Example-based motion generation offers an intuitive way to control animation; however, finding an exact style reference that matches the desired style can be challenging. For instance, a user may want to extract upper-face motion from one example while utilizing lip motion from another. A promising direction for future work is developing a system that allows finer control over style attributes by enabling users to specify and blend motion characteristics from multiple style references to better influence the generation process. Another promising direction is to integrate our VAE-based style space with semantically rich representations, such as the CLIP space proposed in \cite{ma_talkclip_2024}. This integration would enable multi-modal conditioning, allowing the model to generate facial animations informed by both video and text inputs.

% \ep{The proposed framework can also be leveraged to develop robust speech style encoders, transforming facial motion data into compact latent representations, enabling research direction in motion query and classification. offering deeper insights into expressive behaviors and supporting various animation and performance analysis tasks.}
% Secondly, in our paper, we focused problem of style transfer while not fully exploring the design space of diffusion pipelines. One potential avenue of future work would be adapting more compute-efficient inference methods such as DDIM \cite{song_denoising_2022} and utilizing auto-regressive diffusion such as DART \cite{zhao_dart_2024} to build a real-time system. Perhaps we can align our learned VAE space with CLIP embedding to generate labels for speech styles like TalkingCLIP Ma, 2024].

\section{Conclusion}
% We introduced an example-based speech-driven facial animation framework that can accurately transfer the speech mannerisms of a style reference while maintaining accurate lip synchronization, as demonstrated through robust user studies. Our proposed conditioning mechanism, style basis, has been shown to enhance model performance quantitatively across existing benchmarks. This work highlights the potential of example-conditioned facial animation generation, offering a complementary approach to text-conditioned generation and paving the way for the next generation of expressive generative models.
We introduced an example-based speech-driven facial animation framework that captures the distinctive motion characteristics and expressive essence of a reference style while maintaining accurate lip synchronization, as demonstrated through user studies. Our proposed conditioning mechanism, style basis, has been shown to enhance motion accuracy compared to its ablated counterpart, validating its contribution to the overall performance.
This work highlights the potential of example-conditioned facial animation generation, offering a complementary approach to text-conditioned generation. 


%%
%% The acknowledgments section is defined using the "acks" environment
%% (and NOT an unnumbered section). This ensures the proper
%% identification of the section in the article metadata, and the
%% consistent spelling of the heading.
\begin{acks}
We thank Arthur Josi for the assistance working with the SEREP latent model, Emeline Got for rendering the beautiful animations, Marc-André Carbonneau for motivating the idea of using example-based conditioning, and Domenico Tullo for the valuable feedback on our user study design. This project is funded by the Mitacs Accelerate program, IT40568.
\end{acks}

%%
%% The next two lines define the bibliography style to be used, and
%% the bibliography file.
\bibliographystyle{ACM-Reference-Format}
\bibliography{reference_new}

%%
%% If your work has an appendix, this is the place to put it.
% \appendix

\end{document}